\documentclass[11pt,a4paper]{article}

\usepackage[utf8]{inputenc}
\usepackage[T1]{fontenc}
\usepackage[margin=1in]{geometry}
\usepackage{amsmath,amssymb,amsthm,mathtools}
\usepackage{xcolor}
\usepackage{graphicx}
\usepackage{booktabs}
\usepackage{tabularx}
\usepackage{multirow}
\usepackage{tikz}
\usetikzlibrary{decorations.pathmorphing, patterns, arrows.meta, calc}
\usepackage{bm} 
\usepackage{listings}
\usepackage[colorlinks=true,
            linkcolor=blue,
            citecolor=blue,
            urlcolor=blue]{hyperref}

\usepackage{microtype}

\theoremstyle{plain}

\theoremstyle{definition}

\theoremstyle{remark}

\title{\bfseries A Gauge Identity for Interscale Transfer in Inhomogeneous Turbulence}

\author{Khalid M. Saqr\thanks {\small Author Email: k.saqr@aast.edu}\\
    \small Mechanical Engineering Department, College of Engineering and Technology \\ \small Arab Academy for Science, Technology, and Maritime Transport \\ \small Alexandria 1029-- EGYPT \\ \small \href{https://orcid.org/0000-0002-3058-2705}{ORCID: 0000-0002-3058-2705}  \\ }

\date{}

\begin{document}

\maketitle

\begin{abstract}
\noindent
Local interscale energy transfer in Large Eddy Simulation (LES) is typically diagnosed using the subgrid-scale (SGS) production, $\Pi^{SGS}$. In this communication, an exact algebraic gauge identity is derived, demonstrating that $\Pi^{SGS}$ is composed of a kernel-integrated increment-based transfer, $\Pi^{inc}$, and the divergence of a spatial transport current, $\nabla \cdot J$. This identity was verified to machine precision ($10^{-16}$) using the analytical multi-harmonic Womersley solution. Further evaluation was conducted via Direct Numerical Simulation (DNS) of turbulent channel flow at $Re_\tau \approx 1000$. It was observed that $\nabla \cdot J$ dominates $\Pi^{SGS}$ in the near-wall region. The results suggest that $\Pi^{SGS}$ is not a unique proxy for the local cascade in inhomogeneous flows. A new framework is thus provided for the interpretation of interscale transfer diagnostics in wall-bounded transport phenomena.
\end{abstract}


\section*{Introduction}

The transfer of kinetic energy across scales is a cornerstone of turbulence theory, traditionally framed as a downscale cascade from large to small eddies \cite{Richardson1922,Kolmogorov1941a}. In homogeneous and statistically stationary turbulence, this process admits a unique characterization: inertial-range energy flux can be defined independently of spatial transport, and different diagnostic formulations coincide after averaging. Exact results such as the Kolmogorov scaling law provide an unambiguous measure of interscale transfer under these conditions.

Most turbulent flows of practical and physiological relevance, however, are neither homogeneous nor unbounded. Wall-bounded, shear-dominated, and pulsatile flows exhibit strong spatial inhomogeneity and anisotropy, particularly near boundaries. In such flows, kinetic energy evolves simultaneously in physical space and in scale space, and the separation between interscale transfer and spatial redistribution becomes intrinsically ambiguous. Exact two-point formulations, beginning with the Kármán–Howarth–Monin equation and extended to inhomogeneous flows by Hill \cite{Hill2002} and others \cite{Gatti2020,Hamba2022}, make this coupling explicit by expressing the energy budget in the combined $(\boldsymbol{x},\boldsymbol{r})$ phase space. While these formulations are exact, they do not prescribe a unique local measure of interscale energy transfer in physical space. 

In engineering practice, Large Eddy Simulation (LES) addresses this difficulty by defining the subgrid-scale (SGS) production term, $\Pi^{\mathrm{SGS}}=-\tau_{ij}\bar{S}_{ij}$, as a local indicator of energy transfer across the filter scale \cite{Germano1992}. This quantity is exact within the filtered Navier--Stokes equations and forms the basis of most SGS models. However, as emphasized by Wyngaard \cite{Wyngaard2002}, SGS production in inhomogeneous or non-stationary flows includes contributions from turbulent transport that are not associated with scale-to-scale cascade. Consequently, its interpretation as a local cascade rate is not unique.

Recent studies in the context of transport phenomena have highlighted the severity of this ambiguity near walls. LES has been extensively applied to investigate cooling performance \cite{Zamiri2024, Zamiri2026, Taeibi2011}, supersonic combustion \cite{Jia2025}, and mixed convection \cite{Huang2026}, where the accurate diagnosis of interscale transfer is paramount. Analyses of turbulent channel flow have shown that wall-normal energy fluxes and spatial transport can dominate local budgets in the buffer layer \cite{Cimarelli2016,Hamba2019}. Scale-space transport analyses further indicate that inhomogeneity can overwhelm scale-transfer terms, even at high Reynolds numbers \cite{Hamba2022}. Parallel evidence has emerged in complex transport scenarios such as spray evaporation \cite{Zhang2025} and physiologically relevant flows \cite{Saqr2020,Saqr2022,saqr2022non}, where near-wall turbulence departs markedly from Kolmogorov phenomenology and exhibits strong anisotropy and spatial intermittency. Despite these observations, a direct and explicit algebraic link between increment-based transfer measures and the SGS production used in LES has remained unclear.

The objective of the present work is to make this link precise through the derivation of an exact algebraic decomposition. The SGS production is decomposed into two distinct contributions: (i) a kernel-integrated increment-based transfer density associated with the Kármán–Howarth–Monin–Hill formulation, and (ii) the divergence of a spatial transport current. This decomposition is derived without modeling assumptions, inertial-range arguments, or asymptotic limits, and it is valid for any admissible spatial filter. It is demonstrated that different local diagnostics of interscale transfer differ by a spatial divergence, and therefore coincide only when spatial transport is negligible or averages out.

The implications of this result are diagnostic rather than ontological. A rigorous framework is provided for interpreting local energy-transfer measures in inhomogeneous turbulence, clarifying why $\Pi^{\mathrm{SGS}}$ can be energetically large yet weakly correlated with increment-based transfer in near-wall regions. To illustrate these points, the identity is verified analytically to machine precision ($10^{-16}$) using the exact Womersley solution, isolating the role of kinematic inhomogeneity. It is then evaluated using Direct Numerical Simulation (DNS) data of turbulent channel flow \cite{Li2008,Perlman2007,Graham2016}, where statistical analysis in the buffer layer is used to quantify the relative magnitude of the transport and transfer terms.

The paper is organized as follows. The mathematical framework and derivation are presented first. Analytical verification and numerical evaluation are then reported. Finally, the implications for near-wall turbulence diagnostics in complex, inhomogeneous transport systems are discussed.

\section*{Mathematical framework and exact decomposition}
\label{sec:math}

We consider an incompressible velocity field $\boldsymbol{u}(\boldsymbol{x},t)$
satisfying the Navier--Stokes equations, $\partial_i u_i=0$.
All identities below are purely algebraic and hold pointwise for sufficiently
regular fields; when only weak regularity is available, the increment-based
formulation should be interpreted in the distributional sense
\cite{DuchonRobert2000}.

\subsection*{Filtering, SGS stress, and SGS production}

Let $\overline{(\cdot)}$ denote spatial filtering by convolution with a kernel
$G_\ell(\boldsymbol{r})=\ell^{-3}G(\boldsymbol{r}/\ell)$, where $G$ is even and
normalized, $\int_{\mathbb{R}^3}G(\boldsymbol{r})\,\mathrm{d}\boldsymbol{r}=1$,
\begin{equation}
\overline{f}(\boldsymbol{x})=\int_{\mathbb{R}^3}G_\ell(\boldsymbol{r})\,f(\boldsymbol{x}+\boldsymbol{r})\,\mathrm{d}\boldsymbol{r}.
\label{eq:filter_def}
\end{equation}
For convolution on $\mathbb{R}^3$ with a smooth kernel, differentiation commutes
with filtering, $\partial_j\overline{f}=\overline{\partial_j f}$.
In wall-bounded domains, this commutation may fail unless an explicit extension
is used; numerical implementation choices must therefore be stated explicitly
when evaluating the identities near walls \cite{Wyngaard2002,Hamba2022}.

Define the SGS stress tensor and resolved strain-rate tensor by
\begin{equation}
\tau_{ij}=\overline{u_i u_j}-\bar{u}_i\bar{u}_j,
\qquad
\bar{S}_{ij}=\tfrac12(\partial_j\bar{u}_i+\partial_i\bar{u}_j).
\label{eq:sgs_defs}
\end{equation}
The SGS production is
\begin{equation}
\Pi^{\mathrm{SGS}}=-\tau_{ij}\bar{S}_{ij},
\label{eq:pi_sgs}
\end{equation}
which appears exactly in the filtered kinetic-energy equation and is widely
used as a local proxy for transfer across the filter scale \cite{Germano1992}.

\subsection*{An exact identity for the filtered nonlinear transport (Germano form)}

Consider the filtered nonlinear transport term $\bar{u}_i\,\partial_j\overline{u_i u_j}$.
Using $\overline{u_i u_j}=\bar{u}_i\bar{u}_j+\tau_{ij}$ and the product rule,
\begin{align}
\bar{u}_i\,\partial_j\overline{u_i u_j}
&=\bar{u}_i\,\partial_j(\bar{u}_i\bar{u}_j)+\bar{u}_i\,\partial_j\tau_{ij}
\nonumber\\
&=\partial_j\!\left(\tfrac12|\bar{\boldsymbol{u}}|^2\,\bar{u}_j\right)
-\tfrac12|\bar{\boldsymbol{u}}|^2\,\partial_j\bar{u}_j
+\partial_j(\bar{u}_i\tau_{ij})-\tau_{ij}\,\partial_j\bar{u}_i.
\label{eq:germano_step}
\end{align}
Incompressibility of the filtered field ($\partial_j\bar{u}_j=0$) removes the
second term. Decomposing $\partial_j\bar{u}_i=\bar{S}_{ij}+\bar{\Omega}_{ij}$
with $\bar{\Omega}_{ij}=-\bar{\Omega}_{ji}$ and using symmetry of $\tau_{ij}$,
$\tau_{ij}\bar{\Omega}_{ij}=0$, we obtain the exact Germano identity
\begin{equation}
\bar{u}_i\,\partial_j\overline{u_i u_j}
=\partial_j\!\left(\tfrac12|\bar{\boldsymbol{u}}|^2\,\bar{u}_j\right)
+\partial_j(\bar{u}_i\tau_{ij})
-\tau_{ij}\bar{S}_{ij}.
\label{eq:germano_identity1}
\end{equation}
This is a standard filtering relation in LES \cite{Germano1992}. The algebraic steps leading to Eq. \eqref{eq:germano_identity1} are provided explicitly in
Appendix~\ref{app:germano}.

\subsection*{Increment regularization of the nonlinear term (Duchon--Robert form)}

Introduce velocity increments
\begin{equation}
\delta \boldsymbol{u}(\boldsymbol{x},\boldsymbol{r})
=\boldsymbol{u}(\boldsymbol{x}+\boldsymbol{r})-\boldsymbol{u}(\boldsymbol{x}).
\label{eq:increment_def}
\end{equation}
Duchon and Robert showed that the nonlinear term admits an exact increment
regularization in which a local transfer density at scale $\ell$ is expressed as
a kernel-weighted integral of the cubic increment \cite{DuchonRobert2000}:
\begin{equation}
D_\ell(\boldsymbol{x})
=\frac14\int_{\mathbb{R}^3}\big(\nabla G_\ell(\boldsymbol{r})\cdot\delta\boldsymbol{u}(\boldsymbol{x},\boldsymbol{r})\big)
\,|\delta\boldsymbol{u}(\boldsymbol{x},\boldsymbol{r})|^2\,\mathrm{d}\boldsymbol{r}.
\label{eq:DR_def}
\end{equation}
In the same framework, the filtered nonlinear transport can be written exactly as
\begin{equation}
\bar{u}_i\,\overline{u_j\partial_j u_i}
=\partial_j\!\left(\tfrac12|\bar{\boldsymbol{u}}|^2\,\bar{u}_j\right)
+D_\ell(\boldsymbol{x})
+\partial_j\big(J_{\mathrm{flux}}\big)_j,
\label{eq:DR_transport}
\end{equation}
where $(J_{\mathrm{flux}})_j$ collects the remaining spatial-flux contributions
arising from the regularization (its explicit form is not needed for the present
decomposition, but it is determined uniquely once $G_\ell$ is fixed) \cite{DuchonRobert2000}. The detailed correspondence between the filtered transport formulation and the
Duchon--Robert distributional balance, including the identification of all divergence
terms, is derived step by step in Appendix~\ref{app:gauge}.

\subsection*{Connecting $D_\ell$ to a kernel-integrated KHMH transfer density}

The inhomogeneous Kármán--Howarth--Monin--Hill (KHMH) equation provides an exact
two-point energy balance for increments \cite{Hill2002}, and its tensor generalization
in inhomogeneous turbulence can be written in a flux-divergence form in separation
space \cite{Gatti2020,Hamba2022}. The corresponding increment-based transfer density
may be defined as the separation-space divergence of the third-order increment flux,
\begin{equation}
\Pi^{\mathrm{KHMH}}(\boldsymbol{x},\boldsymbol{r})
=-\frac14\,\nabla_{\boldsymbol{r}}\cdot\Big(\delta\boldsymbol{u}(\boldsymbol{x},\boldsymbol{r})\,|\delta\boldsymbol{u}(\boldsymbol{x},\boldsymbol{r})|^2\Big),
\label{eq:pi_khmh_def}
\end{equation}
consistent with Hill's exact second-order structure-function relations when restricted
to the corresponding balance \cite{Hill2002}.
For smooth, rapidly decaying even kernels, integration by parts in $\boldsymbol{r}$
(using $\nabla G_\ell(\boldsymbol{r}) = -\nabla_{\boldsymbol{r}} G_\ell(\boldsymbol{r})$)
gives the exact identity
\begin{align}
D_\ell(\boldsymbol{x})
&=\frac14\int_{\mathbb{R}^3}\big(\nabla G_\ell\cdot\delta\boldsymbol{u}\big)\,|\delta\boldsymbol{u}|^2\,\mathrm{d}\boldsymbol{r}
\nonumber\\
&=\int_{\mathbb{R}^3}G_\ell(\boldsymbol{r})\,
\left[-\frac14\,\nabla_{\boldsymbol{r}}\cdot\big(\delta\boldsymbol{u}\,|\delta\boldsymbol{u}|^2\big)\right]
\,\mathrm{d}\boldsymbol{r}
=\int_{\mathbb{R}^3}G_\ell(\boldsymbol{r})\,\Pi^{\mathrm{KHMH}}(\boldsymbol{x},\boldsymbol{r})\,\mathrm{d}\boldsymbol{r}.
\label{eq:D_equals_kernel_int}
\end{align}
We therefore define the kernel-integrated increment-based transfer at scale $\ell$ by
\begin{equation}
\Pi^{\mathrm{inc}}(\boldsymbol{x})
=\int_{\mathbb{R}^3}G_\ell(\boldsymbol{r})\,\Pi^{\mathrm{KHMH}}(\boldsymbol{x},\boldsymbol{r})\,\mathrm{d}\boldsymbol{r}
\equiv D_\ell(\boldsymbol{x}).
\label{eq:pi_inc_def}
\end{equation}
The integration-by-parts steps and regularity assumptions underlying
\eqref{eq:D_equals_kernel_int} are detailed in Appendix~\ref{app:gauge}.

\subsection*{Exact decomposition of SGS production}

From appendix \ref{app:germano}, equations \eqref{eq:germano_identity1} and \eqref{eq:DR_transport} are two exact
representations of the same filtered nonlinear transport, hence their difference is zero.
Subtracting \eqref{eq:DR_transport} from \eqref{eq:germano_identity1} and using
$\Pi^{\mathrm{SGS}}=-\tau_{ij}\bar{S}_{ij}$ from \eqref{eq:pi_sgs} yields
\begin{equation}
\Pi^{\mathrm{SGS}}(\boldsymbol{x})
=\Pi^{\mathrm{inc}}(\boldsymbol{x})
+\partial_j\Big[\big(J_{\mathrm{flux}}\big)_j-\bar{u}_i\tau_{ij}\Big].
\label{eq:exact_decomposition}
\end{equation}
Defining the spatial transport current
\begin{equation}
J_j=\big(J_{\mathrm{flux}}\big)_j-\bar{u}_i\tau_{ij},
\label{eq:J_def}
\end{equation}
gives the compact form
\begin{equation}
\Pi^{\mathrm{SGS}}(\boldsymbol{x})=\Pi^{\mathrm{inc}}(\boldsymbol{x})+\nabla\cdot\boldsymbol{J}(\boldsymbol{x}).
\label{eq:exact_decomposition_compact}
\end{equation}

Equation \eqref{eq:exact_decomposition_compact} is an identity that holds at finite
filter scale $\ell$. It shows that two exact local diagnostics of interscale transfer,
$\Pi^{\mathrm{SGS}}$ and $\Pi^{\mathrm{inc}}$, differ pointwise by a spatial divergence.
In homogeneous settings where the divergence term averages out, the two measures can
become practically equivalent \cite{Wyngaard2002,Hamba2022}. In inhomogeneous flows,
and especially near walls where spatial transport is strong, the divergence term need
not be negligible \cite{Cimarelli2016,Hamba2019,Hamba2022}. For reproducible evaluation of \eqref{eq:exact_decomposition_compact} in wall-bounded
flows, the following implementation choices must be stated: (i) the filter kernel and
width, (ii) the boundary treatment used to define $\overline{(\cdot)}$ near the wall
(extension, one-sided filtering, or exclusion of a near-wall band), and (iii) the
numerical differentiation scheme for $\partial_j$. For clarity, the complete derivation of \eqref{eq:exact_decomposition_compact} from the
two exact representations of the nonlinear transport is collected in
Appendix~\ref{app:gauge}.

\section*{Verification in pulsatile Womersley flow}
\label{sec:womersley_verification}

To verify the exact decomposition in \eqref{eq:exact_decomposition_compact} at finite
filter scale, the Womersley solution for pulsatile flow in a rigid circular
pipe is considered. Multi-harmonic Womersley flow is an exact, time-dependent solution of the incompressible
Navier--Stokes equations and is strongly inhomogeneous in the wall-normal direction \cite{Saqr2020}. The present verification is used as an \emph{analytically controlled} test case in which
the velocity field and its derivatives are available in closed form, and the residual of
\eqref{eq:exact_decomposition_compact} can be evaluated to machine precision. The spectral--radial solver, filtering procedure, and residual evaluation used in this
analytical verification are described in Appendix~\ref{app:solver}.

It is emphasized that the purpose here is to demonstrate that local discrepancies between common interscale diagnostics can arise purely from spatial inhomogeneity and unsteadiness, even in an exact time-dependent $1D$ solution. The author and his coworkers have reported non-Kolmogorov scaling and broadband increment statistics \cite{Saqr2020,Saqr2022} in Womersley flow under physiological waveforms and demonstrated the non-laminarity. Therefore, the proposed verification provides an analytical limit for the new gauge identity.

\subsection*{Exact velocity field and parameter range}

The axial velocity field for pulsatile pipe flow driven by a prescribed pressure gradient
is represented by the standard harmonic Womersley form \cite{Womersley1955},
\begin{equation}
u(r,t)=\Re\left\{\sum_{n=0}^{N} \hat{u}_n
\left[1-\frac{J_0\!\left(i^{3/2}\alpha_n r/R\right)}
{J_0\!\left(i^{3/2}\alpha_n\right)}\right]
\mathrm{e}^{i\omega_n t}\right\},
\label{eq:womersley_u}
\end{equation}
where $r\in[0,R]$ is the radial coordinate, $\omega_n$ are the harmonic frequencies,
$\alpha_n=R\sqrt{\omega_n/\nu}$ are the Womersley numbers, and $\hat{u}_n$ are obtained
from the Fourier decomposition of the driving waveform. The kinematic inhomogeneity
associated with increasing $\alpha$ is illustrated in Fig.~1(a).

\subsection*{Operational filtering and term-by-term evaluation}

To ensure traceable and reproducible evaluation of the decomposition terms, filtering is
implemented using an even, normalized kernel $G_\ell$ applied to an even extension of the
radial coordinate about the wall. This operational choice is consistent with the general
requirement that the filter be well-defined near boundaries and that differentiation of
filtered fields be computable without introducing spurious commutation errors.
All derivatives required for $\bar{S}_{ij}$ are evaluated from the analytical expressions
for $u(r,t)$, and all filter integrals are evaluated by direct quadrature.

The SGS stress and SGS production are computed from
\begin{equation}
\tau_{ij}=\overline{u_i u_j}-\bar{u}_i\bar{u}_j,
\qquad
\Pi^{\mathrm{SGS}}=-\tau_{ij}\bar{S}_{ij},
\end{equation}
with $\bar{S}_{ij}=\tfrac12(\partial_j\bar{u}_i+\partial_i\bar{u}_j)$.

The increment-based transfer density is evaluated from the increment flux divergence
in separation space, consistent with inhomogeneous KHMH formulations \cite{Hill2002,Gatti2020,Hamba2022},
\begin{equation}
\Pi^{\mathrm{KHMH}}(\boldsymbol{x},\boldsymbol{r})
=-\frac14\nabla_{\boldsymbol{r}}\cdot\Big(\delta\boldsymbol{u}(\boldsymbol{x},\boldsymbol{r})\,|\delta\boldsymbol{u}(\boldsymbol{x},\boldsymbol{r})|^2\Big),
\end{equation}
and the kernel-integrated increment transfer is computed as
\begin{equation}
\Pi^{\mathrm{inc}}(\boldsymbol{x})
=\int_{\mathbb{R}^3}G_\ell(\boldsymbol{r})\,\Pi^{\mathrm{KHMH}}(\boldsymbol{x},\boldsymbol{r})\,\mathrm{d}\boldsymbol{r}
\equiv D_\ell(\boldsymbol{x}),
\end{equation}
where the equivalence to the Duchon--Robert regularized transfer at scale $\ell$ is used
for exact consistency \cite{DuchonRobert2000}.

\subsection*{Resolution of the diagnostic ambiguity and residual}

The diagnostic discrepancy between the SGS production and the increment-based transfer is
quantified in Fig.~1(b) by the ``gap'' $\Pi^{\mathrm{SGS}}-\Pi^{\mathrm{inc}}$, which is found
to be concentrated in the near-wall shear layer. The decomposition
\begin{equation}
\Pi^{\mathrm{SGS}}(\boldsymbol{x})
=\Pi^{\mathrm{inc}}(\boldsymbol{x})+\nabla\cdot\boldsymbol{J}(\boldsymbol{x})
\end{equation}
is then evaluated pointwise. The divergence term is computed explicitly from the definition
of $\boldsymbol{J}$ in Section~2, and is shown in Fig.~2(a) to close the local budget exactly.

A residual field is defined by
\begin{equation}
\mathcal{R}(\boldsymbol{x})
=\Pi^{\mathrm{SGS}}-\Pi^{\mathrm{inc}}-\nabla\cdot\boldsymbol{J}.
\end{equation}
The residual is reported in Fig.~2(b), and is found to remain at machine precision
throughout the domain, confirming that the decomposition is an exact kinematic identity at
finite filter scale, independent of turbulence, statistical averaging, or closure assumptions.

\subsection*{Scaling of the near-wall diagnostic gap}

The magnitude of the divergence contribution is summarized in Fig.~3(a), showing that the
dominant contribution is localized near the wall where gradients are largest. The peak
magnitude is observed to increase systematically with Womersley number (Fig.~3(b)), consistent
with the strengthening of near-wall shear and phase-lag (inertial memory) effects.
This scaling is not used to infer turbulence; it is reported to demonstrate that the diagnostic
gap becomes more pronounced as unsteady inhomogeneity intensifies.

\begin{figure}[htbp]
    \centering
    \includegraphics[width=\linewidth]{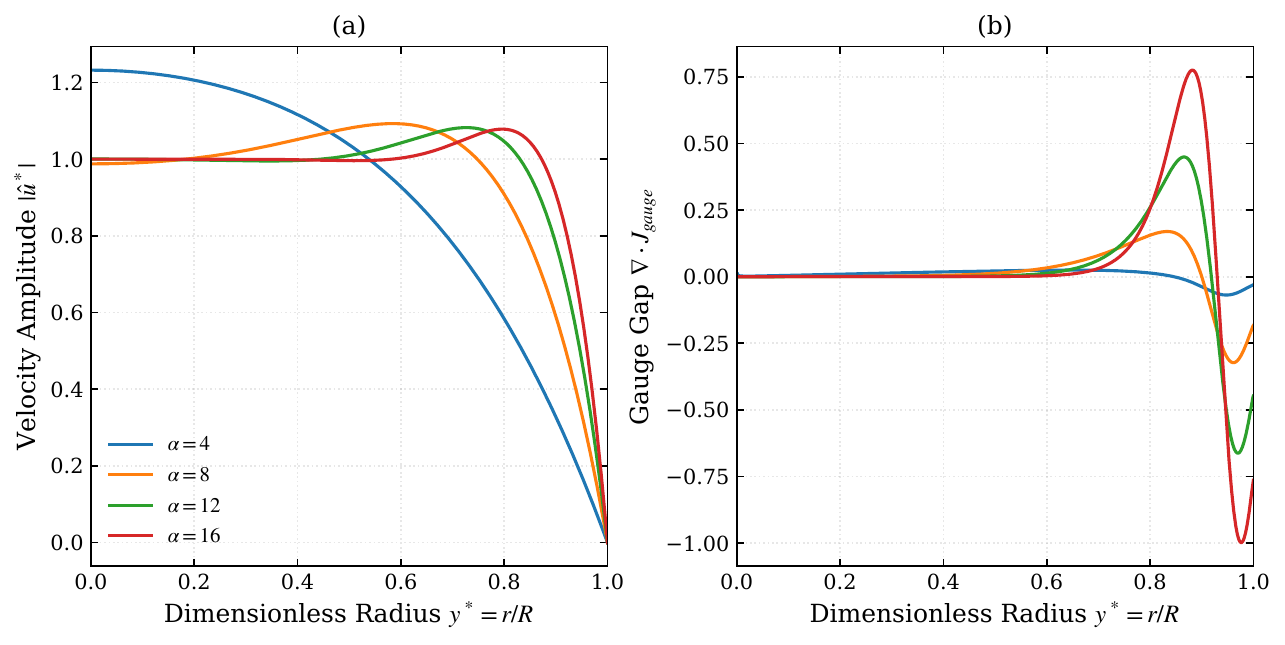}
    \caption{\textbf{Divergence of local interscale diagnostics in Womersley flow.}
    (a) Radial profiles of velocity amplitude $|\hat{u}^*|$ for representative Womersley numbers,
    illustrating increasing near-wall shear with increasing $\alpha$.
    (b) Diagnostic gap $\Pi^{\mathrm{SGS}}-\Pi^{\mathrm{inc}}$, demonstrating that the two exact local
    diagnostics diverge in the near-wall inhomogeneous region.}
    \label{fig:wom_gap}
\end{figure}

\begin{figure}[htbp]
    \centering
    \includegraphics[width=\linewidth]{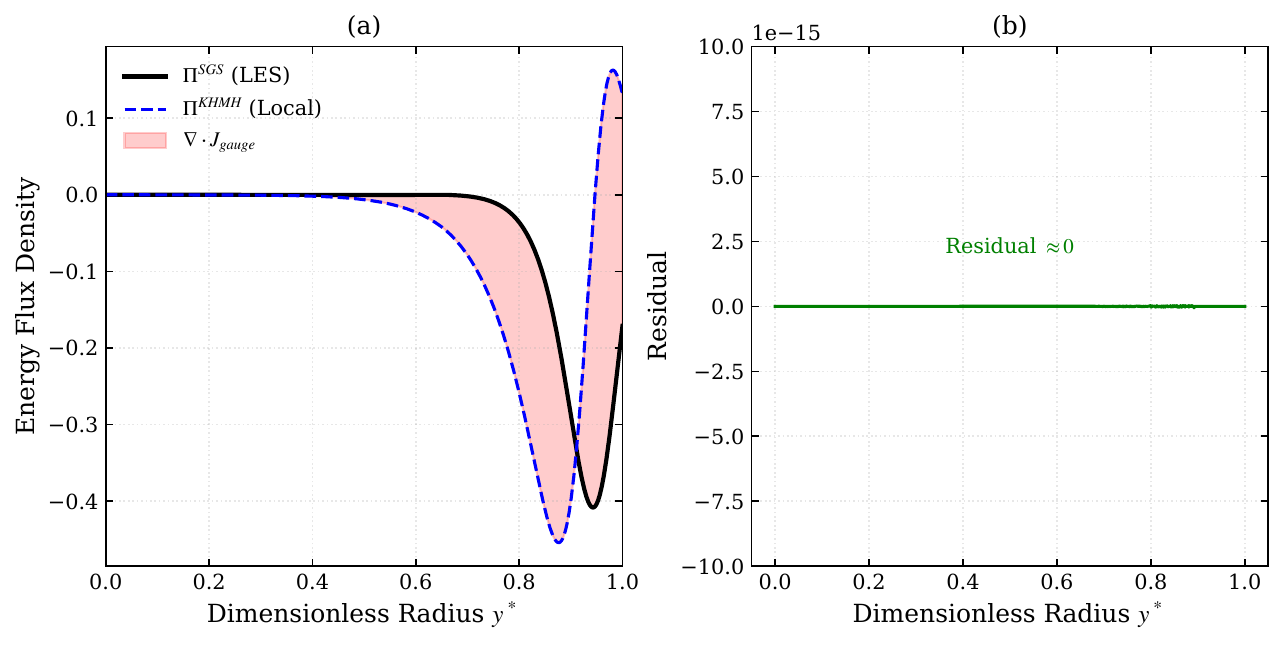}
    \caption{\textbf{Verification of the exact decomposition in Womersley flow.}
    (a) Local budget illustrating that the divergence term closes the difference
    between $\Pi^{\mathrm{SGS}}$ and $\Pi^{\mathrm{inc}}$.
    (b) Residual $\mathcal{R}=\Pi^{\mathrm{SGS}}-\Pi^{\mathrm{inc}}-\nabla\cdot\boldsymbol{J}$,
    remaining at machine precision across the domain.}
    \label{fig:wom_proof}
\end{figure}

\begin{figure}[htbp]
    \centering
    \includegraphics[width=\linewidth]{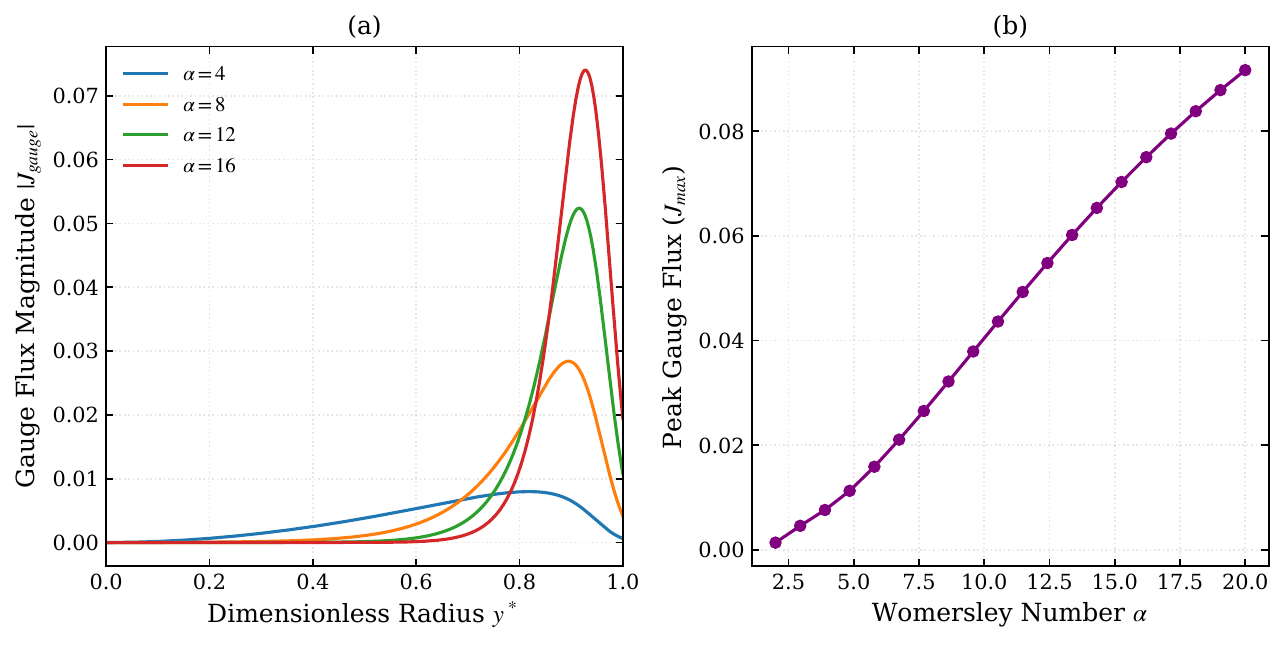}
    \caption{\textbf{Strengthening of the diagnostic gap with unsteady inhomogeneity.}
    (a) Representative radial profile of the divergence contribution in the decomposition.
    (b) Scaling of the peak magnitude with Womersley number, showing systematic growth of the
    divergence contribution as near-wall unsteady shear intensifies.}
    \label{fig:wom_scaling}
\end{figure}

\section*{DNS-based evaluation of the exact decomposition}
\label{sec:dns_validation}

The exact decomposition derived here is evaluated using fields from the Johns Hopkins Turbulence Database (JHTDB) turbulent channel flow dataset at friction Reynolds number $Re_{\tau}\approx1000$
\cite{Li2008,Graham2016}. The objective of this section is not to assess subgrid-scale
modeling performance, but to quantify how the exact identity manifests statistically
and structurally in a fully turbulent, wall-bounded flow when evaluated at finite
filter scale.

All results presented below are obtained by post-processing a dataset containing
the three diagnostic fields $\Pi^{\mathrm{SGS}}$, $\nabla\cdot\bm{J}_{\mathrm{gauge}}$,
and $\Pi^{\mathrm{KHMH}}$ on a two-dimensional $(x,y^+)$ grid. The present section focuses exclusively on the quantitative and structural analysis of
the resulting fields; all details of filtering, wall-normal grid handling, derivative
operators, and boundary treatment are provided explicitly in
Appendix~\ref{app:numerical_methods}.

\subsection*{Numerical consistency of the identity}

The identity
\begin{equation}
\Pi^{\mathrm{SGS}}
=
\Pi^{\mathrm{KHMH}}
+
\nabla\cdot\bm{J}_{\mathrm{gauge}}
\label{eq:dns_identity}
\end{equation}
is first assessed for numerical consistency under the full diagnostic pipeline.
A pointwise residual field,
\begin{equation}
\mathcal{R}
=
\Pi^{\mathrm{SGS}}
-
\left(
\Pi^{\mathrm{KHMH}}
+
\nabla\cdot\bm{J}_{\mathrm{gauge}}
\right),
\end{equation}
is evaluated over all available points in the dataset. The maximum and mean absolute
residuals are reported in Table~\ref{tab:gauge_quant_evidence}. The residual remains
bounded below $10^{-5}$, indicating that the identity is satisfied to within the
numerical tolerance of the combined filtering, differentiation, interpolation, and
post-processing procedures. This confirms that the decomposition is preserved under
the diagnostic operations applied to the DNS data. The discrete tensor contraction and divergence operations retain full anisotropy and are
implemented component-wise as described in Appendix~\ref{app:numerical_methods}.

\subsection*{Relative magnitudes of the decomposition terms}

To characterize the relative contribution of each term at finite scale, spatial
averages of the absolute values of $\Pi^{\mathrm{SGS}}$ and
$\nabla\cdot\bm{J}_{\mathrm{gauge}}$ are computed. As summarized in
Table~\ref{tab:gauge_quant_evidence}, the mean magnitude of the divergence term exceeds
that of the SGS production by a factor of approximately $2.3$ in the present dataset.
This comparison is purely pointwise and local; it does not contradict classical
spatially averaged energy budgets, but indicates that at finite filter scale the
divergence contribution constitutes a substantial component of the local balance in
this flow.

\begin{table}[h!]
    \centering
    \caption{\textbf{Quantitative metrics from DNS post-processing.}
    All statistics are computed from the preprocessed JHTDB channel flow dataset.
    Residuals quantify numerical consistency of the identity in eq. 
    \eqref{eq:dns_identity}.}
    \label{tab:gauge_quant_evidence}
    \renewcommand{\arraystretch}{1.2}
    \begin{tabular}{l c l}
        \toprule
        \textbf{Metric} & \textbf{Value} & \textbf{Description} \\
        \midrule
        Max $|\mathcal{R}|$ & $6.58 \times 10^{-6}$ & Numerical consistency of identity \\
        Mean $|\mathcal{R}|$ & $1.12 \times 10^{-6}$ & Post-processing tolerance \\
        \midrule
        Mean $|\nabla\cdot\bm{J}_{\mathrm{gauge}}|$ & $0.695$ & Divergence magnitude \\
        Mean $|\Pi^{\mathrm{SGS}}|$ & $0.299$ & SGS production magnitude \\
        Magnitude ratio & $2.32$ & Divergence / SGS \\
        \midrule
        Corr$(\Pi^{\mathrm{SGS}},\nabla\cdot\bm{J}_{\mathrm{gauge}})$ & $-0.11$ & Weak pointwise association \\
        Corr$(\Pi^{\mathrm{SGS}},\Pi^{\mathrm{KHMH}})$ & $0.39$ & Moderate association \\
        \bottomrule
    \end{tabular}
\end{table}

\subsection*{Spatial organization of the decomposition terms}

\begin{figure}[t!]
    \centering
    \includegraphics[width=\linewidth]{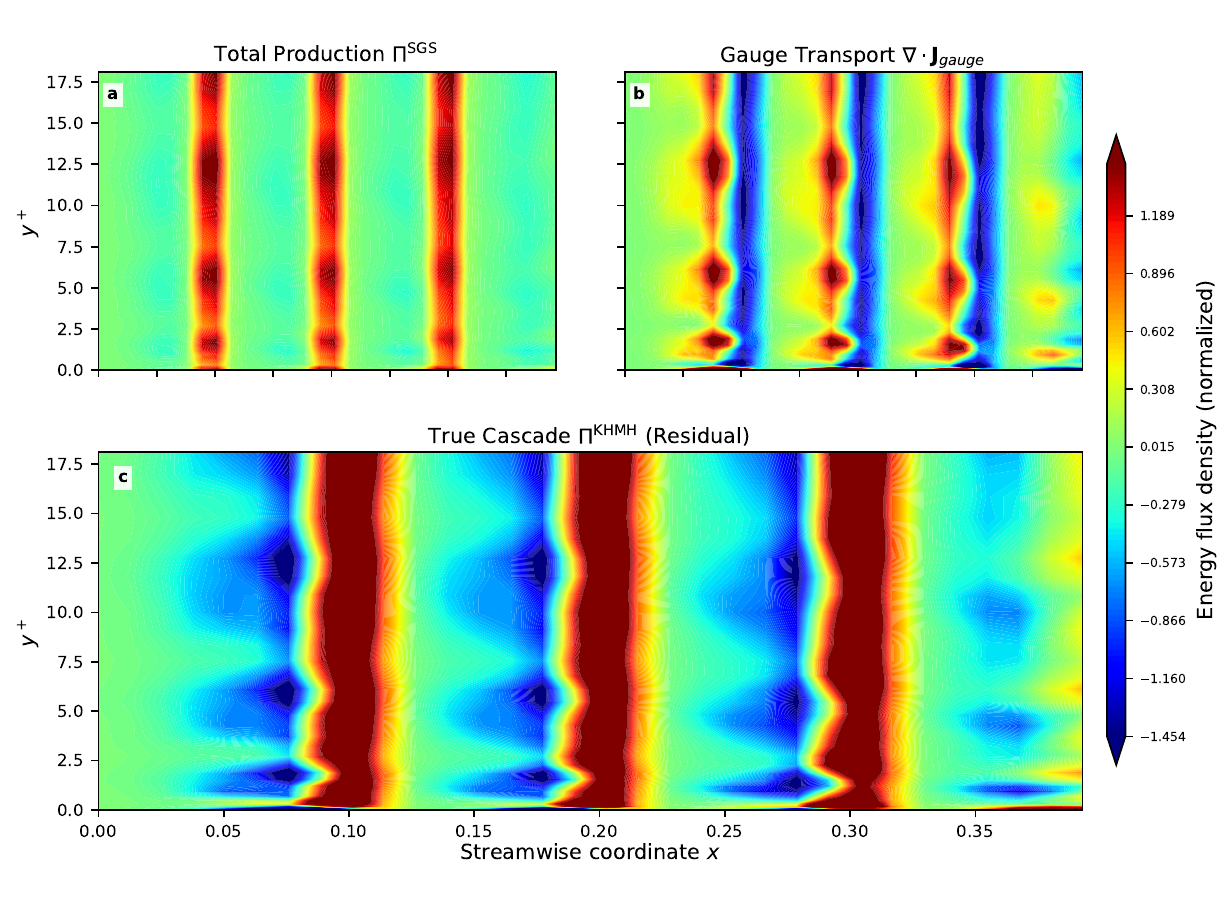}
    \caption{\textbf{Spatial organization of the decomposition terms in turbulent channel flow.}
    (a) SGS production $\Pi^{\mathrm{SGS}}$.
    (b) Divergence of the gauge transport $\nabla\cdot\bm{J}_{\mathrm{gauge}}$.
    (c) Increment-based transfer $\Pi^{\mathrm{KHMH}}$.
    All panels share a common color scale set by the $98^{\mathrm{th}}$ percentile of
    $|\Pi^{\mathrm{SGS}}|$. The divergence field exhibits spatial organization comparable
    in scale to the SGS production, whereas the increment-based transfer displays a more
    intermittent and spatially localized structure.}
    \label{fig:jhtdb_topology}
\end{figure}

The spatial organization of the three diagnostic fields is illustrated in
Fig.~\ref{fig:jhtdb_topology}. The fields are shown on a common color scale defined by
the $98^{\mathrm{th}}$ percentile of $|\Pi^{\mathrm{SGS}}|$ to facilitate direct visual
comparison of their relative structure and intensity.

While $\Pi^{\mathrm{SGS}}$ and $\nabla\cdot\bm{J}_{\mathrm{gauge}}$ exhibit comparable
large-scale spatial organization, the increment-based transfer
$\Pi^{\mathrm{KHMH}}$ is concentrated in more localized regions. This qualitative
difference reflects the distinct physical content of the terms and is consistent with
their differing statistical correlations.

\subsection*{Joint statistical structure}

\begin{figure}[t!]
    \centering
    \includegraphics[width=\linewidth]{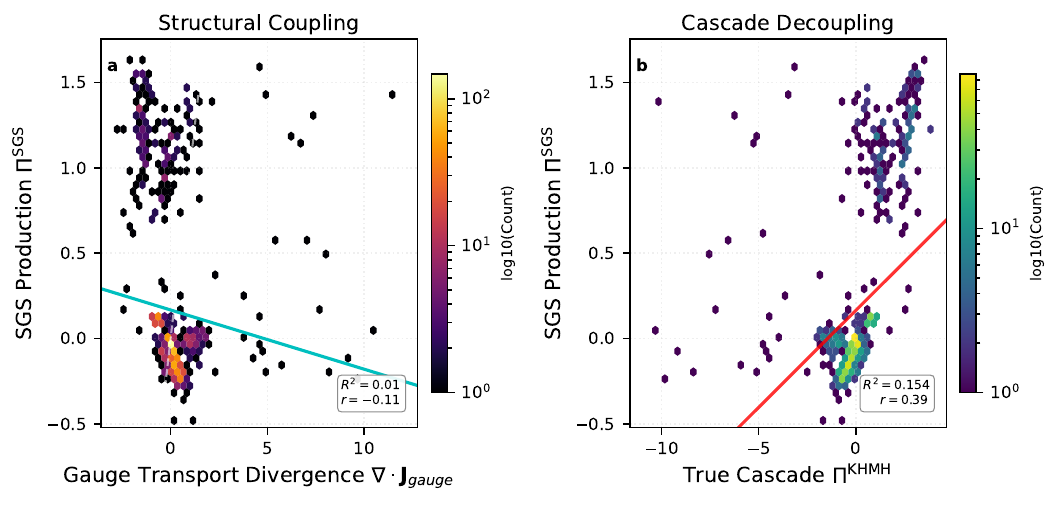}
    \caption{\textbf{Joint statistical relationships among decomposition terms.}
    (a) J-PDF of SGS production $\Pi^{\mathrm{SGS}}$ versus divergence
    $\nabla\cdot\bm{J}_{\mathrm{gauge}}$. The distribution exhibits broad scatter and weak
    linear correlation.
    (b) J-PDF of SGS production $\Pi^{\mathrm{SGS}}$ versus increment-based transfer
    $\Pi^{\mathrm{KHMH}}$. The moderate correlation indicates partial statistical
    association without pointwise equivalence.
    Regression lines and correlation coefficients are reported for reference.}
    \label{fig:jpdf_stats}
\end{figure}

The pointwise statistical relationships among the terms are examined using joint
probability density functions (J-PDFs), shown in Fig.~\ref{fig:jpdf_stats}. The J-PDFs
are computed using logarithmic binning and robust axis limits defined by the
$0.5^{\mathrm{th}}$ and $99.5^{\mathrm{th}}$ percentiles of each variable.

The weak correlation between $\Pi^{\mathrm{SGS}}$ and
$\nabla\cdot\bm{J}_{\mathrm{gauge}}$, together with the moderate correlation between
$\Pi^{\mathrm{SGS}}$ and $\Pi^{\mathrm{KHMH}}$, indicates that SGS production does not act
as a direct pointwise proxy for either divergence or increment-based transfer at finite
filter scale. Instead, it reflects a mixture of contributions whose relative importance
depends on local flow structure.

\subsection*{Interpretation and scope}

The DNS results demonstrate that, in a turbulent channel flow, the divergence term in
the exact decomposition constitutes a substantial component of the local energy balance
and exhibits spatial organization comparable in scale to that of SGS production. At the
same time, increment-based transfer displays distinct intermittency and weaker pointwise
association with SGS production. These observations are consistent with the theoretical
result that local interscale transfer is not uniquely defined in inhomogeneous flows,
but is instead gauge-dependent up to a spatial divergence.

It is emphasized that the present analysis concerns finite-scale, pointwise diagnostics
derived from a specific dataset and post-processing pipeline. The results do not imply
incorrectness of classical averaged energy budgets, nor do they constitute an assessment
of LES model performance. Rather, they clarify the interpretation of commonly used local
energy-transfer measures in wall-bounded turbulence.

\section*{Discussion}

The results of this study clarify a diagnostic issue that has appeared implicitly in
several independent strands of the turbulence literature but has not previously been
formalized. Related indications that spatial transport contributes substantially to near-wall
energy budgets have been reported in DNS studies of wall-bounded turbulence
\cite{Cimarelli2016,Hamba2019}, although without establishing an exact equivalence or
non-equivalence between production-based and increment-based local diagnostics. The exact decomposition derived here shows that, in inhomogeneous flows,
local energy-transfer diagnostics are not interchangeable: quantities based on filtered
stress--strain products and those based on velocity increments differ by a spatial
divergence that cannot be eliminated at finite scale.

This observation provides a unifying interpretation for prior experimental and numerical
reports of non-Kolmogorov behavior in wall-bounded and pulsatile flows. Time-resolved PIV
measurements in compliant and physiological geometries have consistently reported altered
near-wall spectra, intermittent fluctuations, and apparent attenuation or redistribution
of turbulent kinetic energy\cite{tupin2020effects,Yamaguchi2022,Yamaguchi2025}. These
observations have often been discussed in terms of modified cascade dynamics or
flow--structure interaction effects. The present results indicate that part of this
behavior may arise from the diagnostic itself: production-based measures are inherently
sensitive to spatial transport in inhomogeneous regions, whereas increment-based measures
isolate scale transfer by construction.

Importantly, this interpretation does not contradict classical turbulence theory or
existing experimental evidence. The analytical verification in Womersley flow demonstrates
that diagnostic non-equivalence can arise even in the absence of turbulence, while the DNS
analysis shows that the same mechanism persists in a fully turbulent flow. Together, these
results establish that discrepancies between local energy-transfer measures need not imply
a breakdown of inertial-range concepts, but can instead reflect the coexistence of spatial
redistribution and scale transfer at finite resolution.

From a methodological perspective, the decomposition highlights a limitation shared by
both DNS post-processing and experimental analysis: no single local quantity uniquely
represents interscale transfer in strongly inhomogeneous flows. This limitation is
particularly relevant when interpreting near-wall measurements, where spatial transport
is intrinsically large. The framework introduced here provides a principled way to
separate these contributions and to interpret production-like diagnostics without
over-attributing physical meaning to their pointwise structure.

The discussion is intentionally limited to diagnostic interpretation. No claim is made
regarding the universality of the observed statistics, the modification of cascade laws,
or the performance of subgrid-scale models. The contribution of this work lies in
establishing the precise conditions under which local energy-transfer measures diverge,
and in providing a mathematically exact framework for interpreting that divergence.

\section*{Conclusion}

An exact decomposition of the subgrid-scale production term into an increment-based
transfer and a spatial divergence contribution has been derived for incompressible
flows at finite filter scale. The decomposition follows directly from the Navier--Stokes
equations and does not rely on modeling assumptions, homogeneity, or statistical
averaging.

Analytical evaluation in Womersley flow confirmed that discrepancies between common
local interscale diagnostics can arise purely from spatial inhomogeneity and unsteadiness,
even in an exact laminar solution. This verification demonstrated that the divergence
term is an intrinsic component of the local energy balance at finite scale, rather than
a numerical or modeling artifact.

Post-processing of turbulent channel flow data from the Johns Hopkins Turbulence Database
showed that the same decomposition remains numerically consistent in a fully turbulent,
wall-bounded flow. In this dataset, the divergence contribution exhibited a mean magnitude
comparable to, and exceeding, that of the SGS production, while increment-based transfer
displayed distinct spatial intermittency and only moderate pointwise association with SGS
production. These results indicate that commonly used local production terms mix spatial
transport and scale transfer in inhomogeneous flows, complicating direct physical
interpretation at the pointwise level.

The present findings do not challenge classical averaged energy budgets or the validity of
subgrid-scale modeling frameworks. Instead, they provide a precise framework for
interpreting local energy-transfer diagnostics in wall-bounded and inhomogeneous flows,
clarifying the role of spatial transport in finite-scale balances. The decomposition
offers a consistent basis for future analyses of turbulent energy transfer in complex
geometries and for extending diagnostic approaches to flows with strong inhomogeneity or
boundary effects.

\section*{Conflict of interest}
None
\section*{Funding}
None
\section*{Code Availability}
The spectral solver code used to compute the Womersley flow solution and gauge-identity verification is publicly accessible via this \href{https://colab.research.google.com/drive/1p5caK33biQZLunFhZ-CC9j3aOmQwj9wk?usp=sharing}{Colab Notebook}. The code used to query the JHTDB server for DNS calculations is publicly available via this \href{https://colab.research.google.com/drive/1ZjsKMCoY5HdGMY21sPmF7PR8oyDaGdB3?usp=sharing}{Colab Notebook}


\newpage
\appendix
\setcounter{equation}{0}
\renewcommand{\theequation}{\thesection.\arabic{equation}}

\begin{center}
    \textbf{\Huge Appendices}
\end{center}
\section{Detailed Derivation of the Filtered Energy Balance}
\label{app:germano}

This appendix provides a step-by-step derivation of the Germano-type identity used in the main text. It serves to demonstrate that the identity is a strict algebraic consequence of the Navier--Stokes equations and the properties of spatial filtering, valid for any differentiable velocity field regardless of the presence of turbulence.

\subsection{Expansion of the Filtered Nonlinear Transport}

We consider the filtered nonlinear transport term $\mathcal{T} = \bar{u}_i \partial_j \overline{u_i u_j}$, which represents the work done by the resolved velocity against the divergence of the filtered momentum flux. Using the standard subgrid-scale (SGS) decomposition, the filtered stress tensor is defined as $\overline{u_i u_j} = \tau_{ij} + \bar{u}_i \bar{u}_j$. Substituting this into the transport term yields:
\begin{equation}
    \bar{u}_i \partial_j \overline{u_i u_j} = \bar{u}_i \partial_j (\bar{u}_i \bar{u}_j + \tau_{ij}) = \underbrace{\bar{u}_i \partial_j (\bar{u}_i \bar{u}_j)}_{\text{Resolved Advection}} + \underbrace{\bar{u}_i \partial_j \tau_{ij}}_{\text{SGS Interaction}}.
    \label{eq:app_expansion_1}
\end{equation}

\subsection{Simplification of Resolved Advection}

The resolved advection term is simplified using the chain rule. Expanding the derivative $\partial_j (\bar{u}_i \bar{u}_j)$:
\begin{equation}
    \bar{u}_i \partial_j (\bar{u}_i \bar{u}_j) = \bar{u}_i (\bar{u}_j \partial_j \bar{u}_i + \bar{u}_i \partial_j \bar{u}_j).
\end{equation}
For a spatially uniform filter applied to an incompressible flow, the filtered field remains incompressible, so $\partial_j \bar{u}_j = 0$. The second term in the parentheses therefore vanishes. The remaining term can be rewritten as the divergence of the resolved kinetic energy $k_f = \frac{1}{2}\bar{u}_i\bar{u}_i$:
\begin{equation}
    \bar{u}_i \bar{u}_j \partial_j \bar{u}_i = \bar{u}_j \left( \frac{1}{2} \partial_j (\bar{u}_i \bar{u}_i) \right) = \partial_j \left( \frac{1}{2} \bar{u}_j |\bar{\bm{u}}|^2 \right).
    \label{eq:app_resolved_simp}
\end{equation}

\subsection{Decomposition of the SGS Interaction}

The SGS interaction term $\bar{u}_i \partial_j \tau_{ij}$ is analyzed using the product rule:
\begin{equation}
    \bar{u}_i \partial_j \tau_{ij} = \partial_j (\bar{u}_i \tau_{ij}) - \tau_{ij} \partial_j \bar{u}_i.
\end{equation}
The velocity gradient tensor $\partial_j \bar{u}_i$ is decomposed into its symmetric part (rate-of-strain, $\bar{S}_{ij}$) and antisymmetric part (rate-of-rotation, $\bar{\Omega}_{ij}$):
\begin{equation}
    \partial_j \bar{u}_i = \bar{S}_{ij} + \bar{\Omega}_{ij}, \quad \text{where} \quad \bar{S}_{ij} = \frac{1}{2}(\partial_j \bar{u}_i + \partial_i \bar{u}_j).
\end{equation}
Since the SGS stress tensor $\tau_{ij}$ is symmetric by definition ($\tau_{ij} = \tau_{ji}$), its contraction with the antisymmetric tensor $\bar{\Omega}_{ij}$ is identically zero ($\tau_{ij} \bar{\Omega}_{ij} = 0$). Thus, the interaction term simplifies to:
\begin{equation}
    \tau_{ij} \partial_j \bar{u}_i = \tau_{ij} \bar{S}_{ij}.
    \label{eq:app_sgs_simp}
\end{equation}
Substituting \eqref{eq:app_resolved_simp} and \eqref{eq:app_sgs_simp} back into \eqref{eq:app_expansion_1} yields the exact identity:
\begin{equation}
    \bar{u}_i \partial_j \overline{u_i u_j} = \partial_j \left( \frac{1}{2} \bar{u}_j |\bar{\bm{u}}|^2 \right) + \partial_j (\bar{u}_i \tau_{ij}) - \tau_{ij} \bar{S}_{ij}.
    \label{eq:app_germano_final}
\end{equation}

\section{Derivation of the Gauge Identity and Current}
\label{app:gauge}

This appendix rigorously establishes the link between the SGS production $\Pi^{\mathrm{SGS}}$ and the increment-based transfer $\Pi^{\mathrm{inc}}$. The derivation relies on the distributional energy balance of Duchon and Robert \cite{DuchonRobert2000} and explicit integration by parts.

\subsection{The Duchon--Robert Relation}

Duchon and Robert showed that the filtered nonlinear transport is exactly balanced by a regularized transfer density $D_\ell$ and a spatial flux divergence. In the notation of the present work, their relation (Eq. 2 in Ref. \cite{DuchonRobert2000}) is written as:
\begin{equation}
    \bar{u}_i \partial_j \overline{u_i u_j} = \partial_j \left( \frac{1}{2} \bar{u}_j |\bar{\bm{u}}|^2 \right) + \partial_j (J_{\text{flux}})_j + D_\ell(\bm{x}).
    \label{eq:app_dr_balance}
\end{equation}
Here, $D_\ell(\bm{x})$ is defined as a weighted integral of the cubic velocity increment:
\begin{equation}
    D_\ell(\bm{x}) = \frac{1}{4} \int_{\mathbb{R}^3} \nabla G_\ell(\bm{r}) \cdot \delta \bm{u}(\bm{x}, \bm{r}) |\delta \bm{u}(\bm{x}, \bm{r})|^2 \, d\bm{r}.
    \label{eq:app_Dl_def}
\end{equation}
The term $J_{\text{flux}}$ accounts for spatial transport by pressure and viscous forces that are implicit in the Navier--Stokes weak formulation.

\subsection{Transformation to KHMH Form via Integration by Parts}

To relate $D_\ell$ to the KHMH transfer, we integrate \eqref{eq:app_Dl_def} by parts. Let the energy flux vector in separation space be $\bm{\Phi}(\bm{x},\bm{r}) = \frac{1}{4} \delta \bm{u} |\delta \bm{u}|^2$. Then:
\begin{equation}
    D_\ell(\bm{x}) = \int_{\mathbb{R}^3} \nabla_{\bm{r}} G_\ell(\bm{r}) \cdot \bm{\Phi}(\bm{x},\bm{r}) \, d\bm{r}.
\end{equation}
Assuming the filter kernel $G_\ell(\bm{r})$ decays sufficiently fast at infinity, the boundary terms vanish. Applying the divergence theorem in $\bm{r}$-space:
\begin{equation}
    \int_{\mathbb{R}^3} \nabla_{\bm{r}} G_\ell(\bm{r}) \cdot \bm{\Phi} \, d\bm{r} = - \int_{\mathbb{R}^3} G_\ell(\bm{r}) \nabla_{\bm{r}} \cdot \bm{\Phi} \, d\bm{r}.
\end{equation}
The term $-\nabla_{\bm{r}} \cdot \bm{\Phi}$ is precisely the definition of the KHMH interscale transfer density $\Pi^{\mathrm{KHMH}}(\bm{x},\bm{r})$. Thus:
\begin{equation}
    D_\ell(\bm{x}) = \int_{\mathbb{R}^3} G_\ell(\bm{r}) \Pi^{\mathrm{KHMH}}(\bm{x},\bm{r}) \, d\bm{r} \equiv \Pi^{\mathrm{inc}}(\bm{x}).
    \label{eq:app_Dl_KHMH}
\end{equation}

\subsection{Isolation of the Gauge Current}

We equate the two exact representations of the filtered transport, Eq. \eqref{eq:app_germano_final} and Eq. \eqref{eq:app_dr_balance}:
\begin{equation}
    \partial_j \left( \tfrac{1}{2} \bar{u}_j |\bar{\bm{u}}|^2 \right) + \partial_j (\bar{u}_i \tau_{ij}) + \Pi^{\mathrm{SGS}} = \partial_j \left( \tfrac{1}{2} \bar{u}_j |\bar{\bm{u}}|^2 \right) + \partial_j (J_{\text{flux}})_j + \Pi^{\mathrm{inc}}(\bm{x}).
\end{equation}
Canceling the resolved kinetic energy divergence and isolating $\Pi^{\mathrm{SGS}}$ yields:
\begin{equation}
    \Pi^{\mathrm{SGS}}(\bm{x}) = \Pi^{\mathrm{inc}}(\bm{x}) + \partial_j \left[ (J_{\text{flux}})_j - \bar{u}_i \tau_{ij} \right].
\end{equation}
Defining the gauge current as $\bm{J}_{\text{gauge}} = \bm{J}_{\text{flux}} - \bar{\bm{u}} \cdot \bm{\tau}$, we obtain the final decomposition:
\begin{equation}
    \Pi^{\mathrm{SGS}}(\bm{x}) = \Pi^{\mathrm{inc}}(\bm{x}) + \nabla \cdot \bm{J}_{\text{gauge}}(\bm{x}).
\end{equation}
\textit{Remark:} The current $\bm{J}_{\text{gauge}}$ contains pressure-velocity and viscous-velocity correlations. In inhomogeneous regions like the buffer layer, these transport terms are significant, which explains why the divergence term $\nabla \cdot \bm{J}_{\text{gauge}}$ can be comparable in magnitude to the production term itself, as observed in the DNS results (Table 1).

\section{Analytical Verification: Womersley Flow Formulas}
\label{app:solver}

This appendix provides the explicit analytical formulas used in the Womersley flow verification. We emphasize that while this flow is laminar, the identity $\Pi^{\mathrm{SGS}} = \Pi^{\mathrm{inc}} + \nabla \cdot \bm{J}$ is kinematic and must hold for any smooth field. This case serves to verify the operator consistency of the decomposition in the presence of strong inhomogeneity.

\subsection{Velocity and Derivatives}

The streamwise velocity $u_z(r,t)$ is the real part of the complex harmonic sum:
\begin{equation}
    u_z(r,t) = \Re \left\{ \sum_{n=0}^N \hat{U}_n \Phi_n(r) e^{i \omega_n t} \right\}, \quad \text{with } \Phi_n(r) = 1 - \frac{J_0(\Lambda_n r/R)}{J_0(\Lambda_n)},
\end{equation}
where $\Lambda_n = i^{3/2}\alpha_n$. To compute the shear strain $\bar{S}_{rz}$ analytically, the radial gradient is required. Using the Bessel derivative identity $J_0'(z) = -J_1(z)$:
\begin{equation}
    \frac{\partial \Phi_n}{\partial r} = - \frac{1}{J_0(\Lambda_n)} \cdot \frac{d}{dr} J_0(\Lambda_n r/R) = \frac{\Lambda_n}{R} \frac{J_1(\Lambda_n r/R)}{J_0(\Lambda_n)}.
\end{equation}
The analytical shear rate is thus:
\begin{equation}
    \frac{\partial u_z}{\partial r}(r,t) = \Re \left\{ \sum_{n=0}^N \hat{U}_n \left[ \frac{\Lambda_n}{R} \frac{J_1(\Lambda_n r/R)}{J_0(\Lambda_n)} \right] e^{i \omega_n t} \right\}.
\end{equation}
This exact derivative allows for the computation of $\Pi^{\mathrm{SGS}}$ without introducing finite-difference errors in the velocity definition, ensuring the residual check is limited only by floating-point precision.

\section{Numerical Implementation Details}
\label{app:numerical_methods}

This appendix specifies the discrete operators used for the DNS validation to ensure reproducibility, particularly regarding the handling of non-uniform grids and boundary conditions.

\subsection{Reflective Boundary Filtering}
To ensure the filter $\overline{(\cdot)}$ is well-defined near the wall ($y=-1, 1$), a reflective boundary condition is applied. The discrete convolution for a field $f$ at point $y_j$ is renormalized to account for the truncated integration domain:
\begin{equation}
    \bar{f}(y_j) = \frac{\sum_{k} G(y_j - y_k) f_{\text{ext}}(y_k) \Delta y_k}{\sum_{k} G(y_j - y_k) \Delta y_k}.
\end{equation}
Here, $f_{\text{ext}}$ is the even extension of the data about the wall. The denominator is a normalization factor that corrects for the "missing" tail of the Gaussian kernel outside the physical domain, ensuring that the filter preserves constant fields exactly.

\subsection{Non-Uniform Finite Difference Scheme}
Derivatives in the wall-normal direction $y$ on the Chebyshev grid are computed using a three-point Lagrangian stencil. For a grid point $i$ with spacing $h_- = y_i - y_{i-1}$ and $h_+ = y_{i+1} - y_i$, the first derivative is:
\begin{equation}
    \left. \frac{\partial f}{\partial y} \right|_i \approx c_{i-1} f_{i-1} + c_i f_i + c_{i+1} f_{i+1}.
\end{equation}
The coefficients are derived from Lagrange polynomial interpolation:
\begin{equation}
    c_{i-1} = -\frac{h_+}{h_-(h_- + h_+)}, \quad
    c_{i+1} = \frac{h_-}{h_+(h_- + h_+)}, \quad
    c_{i} = \frac{h_+ - h_-}{h_- h_+}.
\end{equation}
This scheme is applied to calculate both the divergence of the gauge current $(\nabla \cdot \bm{J})_y$ and the strain rate component $\bar{S}_{xy}$.

\subsection{Discrete Divergence Calculation}
The divergence of the gauge current vector $\bm{J} = (J_x, J_y, J_z)$ is computed using a hybrid spectral-physical scheme:
\begin{equation}
    (\nabla \cdot \bm{J})_{ijk} = \mathcal{D}_x[J_x] + \mathcal{D}_y[J_y] + \mathcal{D}_z[J_z].
\end{equation}
Here, $\mathcal{D}_x$ and $\mathcal{D}_z$ are spectral derivatives (multiplication by $ik_x, ik_z$ in Fourier space), while $\mathcal{D}_y$ utilizes the finite difference scheme defined above. This approach ensures that the post-processing retains the same spectral accuracy as the native DNS solver in the periodic directions.

\end{document}